\documentclass[onecollarge,natbib]{svjour2}
\bibpunct{[}{]}{;}{n}{}{,} % to get "[numbered]" references from natbib
\smartqed  % flush right qed marks, e.g. at end of proof
\usepackage{graphicx}
\usepackage{amsmath}
\usepackage{slashed}
\usepackage{amstext}
\usepackage{amssymb}
\usepackage{bm}
\usepackage{float}

\def\bea{\begin{eqnarray}}
\def\eea{\end{eqnarray}}

\newcommand{\sigp}{{\bm \sigma}\cdot{\hat{\bf p}}}

\newcommand{\xip}{{\bm \xi}\cdot{\hat{\bf p}}}

\newcommand{\sigxi}{{\bm \sigma}\cdot{\hat{\bm \xi}}}

\journalname{Few-Body Systems}
\begin{document}

\title{Application of the Covariant Spectator Theory to the study of heavy and heavy-light mesons%\thanks{Grants or other notes
}


\author{Sofia Leit\~ao \and
	Alfred Stadler  \and
          M\,\,T.\,Pe\~na \and
        Elmar P.\,Biernat
}

\authorrunning{ S.\,Leit\~ao,A.\,Stadler, M.\,T.\,Pe\~na, E.\,P.\,Biernat} % if too long for running head

\institute{Sofia Leit\~ao \at
	CFTP, Instituto Superior T\'ecnico, Universidade de Lisboa, Av. Rovisco Pais 1, 1049-001 Lisboa, Portugal
  \email{sofia.leitao@tecnico.ulisboa.pt}
           \and
           A.\,Stadler \at
           Departamento de F\'isica da Universidade de \'Evora, 7000-671 \'Evora, Portugal\\
	\and
	 A.\ Stadler, M.\,T.\,Pe\~na and  Elmar\,P. Biernat \at
	CFTP, Instituto Superior T\'ecnico, Universidade de Lisboa, Av. Rovisco Pais 1, 1049-001 Lisboa, Portugal
}

\date{Received: date / Accepted: date}

\maketitle

\begin{abstract}
As an application of the Covariant Spectator Theory (CST) we calculate the spectrum of heavy-light and heavy-heavy mesons using covariant versions of a linear confining potential, a one-gluon exchange, and a constant interaction. The CST equations possess the correct one-body limit and are therefore well-suited to describe mesons in which one quark is much heavier than the other. We find a good fit to the mass spectrum of heavy-light and heavy-heavy mesons with just three parameters (apart from the quark masses). Remarkably, the fit parameters are nearly unchanged when we fit to experimental pseudoscalar states only or to the whole spectrum. Because pseudoscalar states are insensitive to spin-orbit interactions and do not determine spin-spin interactions separately from central interactions, this result suggests that it is the covariance of the kernel that correctly predicts the spin-dependent quark-antiquark interactions.

\end{abstract}

\section{Introduction}
\label{intro}
At low energies and large distances quarks and gluons interact strongly, and therefore the standard perturbative methods cannot be used for a meaningful description. In order to study the most striking features of the strong interaction, dynamical chiral-symmetry breaking and color confinement, one needs to employ nonperturbative tecniques.

In this work we concentrate on mesons that can be described as strongly-bound states of one quark and one antiquark. The physics of mesons is a very active field of research, especially due to the vast amount of data currently being collected in experimental facilities such as the LHC, BaBaR, Belle, and CLEO. In the near future, exciting results are also expected from Jefferson Lab (GlueX) and FAIR (PANDA).\,Theoretical predictions are therefore important not only to guide the identification of new states---some of them with exotic non-$q\bar{q}$ content \cite{Choi:2003zl}---but also to calculate other observables, such as form factors, decay rates, etc., important for the study of the structure of mesons. 

In the recent years, lattice QCD approaches have made impressive progress, providing us with a large amount of valuable predictions for physical observables~\cite{Kl}. At the same time, non-perturbative continuum methods (for a good review see Ref.~\cite{Brambilla}) have attracted attention with their potential of providing a deeper understanding of QCD in the infrared regime from information that cannot be extracted from lattice data alone. Very recently, Hamiltonian approaches with a phenomenological confinement obtained from light-front holographic QCD~\cite{YangLi} and renormalization-group procedures for effective particles~\cite{Gomez} have been used to study heavy quarkonium. 

Our approach, the Covariant Spectator Theory (CST)~\cite{Gro69}, is close in spirit to the Dyson-Schwinger/ Bethe-Salpeter (DSBS) formalism~\cite{Hilger,Eichmann}. They both aim at a quantum-field-theoretical description where the one- and two-body dynamics are treated self-consistently. Unlike DSBS, CST works directly in Minkowski space. In addition, the two-body CST equation sums the infinite series of all ladder and crossed-ladder exchange diagrams more efficiently than the ladder Bethe-Salpeter equation (BSE), due to important cancellations that occur when the mass of one of the constituent particles becomes large.  It has been proven that, in a scalar theory and in the limit of the heavy mass tending to infinity, these cancellations even become exact, and that the CST equation therefore gives the exact result (for more details, see Ref.~\cite{CSTreview}). 

Further virtues of the CST worth mentioning include:
\begin{itemize}
\item Meson wave functions are given in terms of covariant vertex functions which have simple transformation properties under Lorentz boosts.
\item CST equations are manifestly covariant, but nevertheless require only three-dimensional loop integrations.
\item The two-body CST equation reduces in the one-body limit to the Dirac equation, and in the nonrelativistic limit to the Sch\"odinger equation.
\end{itemize}

This paper is organized as follows: In Section~\ref{sec:2} we introduce the formalism, in Section~\ref{sec:3} we present the results and discussion, and in Section \ref{sec:4} we conclude with a summary and an outlook.

\section{Formalism}
\label{sec:2}

\begin{figure}[h!]
\centering
\includegraphics[width=8cm,clip]{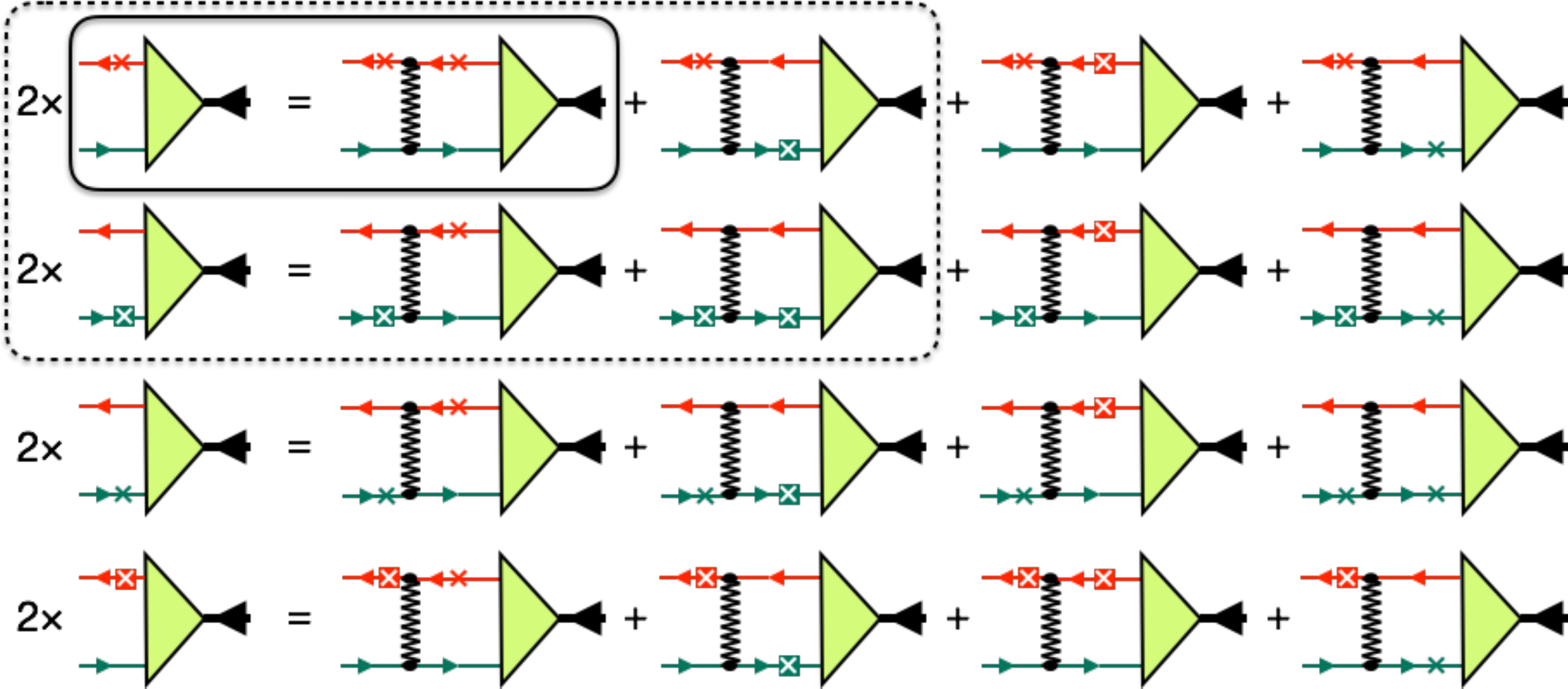}
\caption{The set of the four-channel CST equations (4CSE). The solid rectangle indicates the one-channel CST equation (1CSE) used in this work, the dashed rectangle a two-channel extension with charge-conjugation symmetry. Crosses on quark lines indicate that only the positive-energy pole contribution of the propagator is kept, light crosses in a dark square refer to the negative-energy pole contribution.}
\label{fig}       % Give a unique label
\end{figure}

In order to derive the CST set of equations we start with the BSE for the quark-antiquark vertex function $\Gamma_{BS}(p_1,p_2)$  with an irreducible interaction kernel ${\cal V}(p,k;P)$, where $P$ is the total four-momentum, and $p$ and $k$ are the external and internal relative four-momenta, respectively. The BSE is given by
\begin{equation}
\Gamma_{BS}(p_1,p_2)= i\int\frac{ {d}^4k}{(2\pi)^4}\,{\cal V}(p,k;P) 
S_1({k}_1)\,\Gamma_{BS}(k_1,k_2)\,S_2(k_2)\,,
\label{eq:BS}
\end{equation}
where $S_i(k_i)$ is the dressed quark propagator depending on the individual four-momentum $k_i$ of quark $i$. In the CST, the heavier quark, say quark 1 with mass $m_1$, is on-mass-shell. This yields the CST equation for the vertex function $\Gamma_{1CS}$, where \lq\lq 1CS" or \lq\lq 1CSE" stands for one-channel spectator equation \cite{CST:2014}. More specifically, the 1CSE results from the BSE by keeping in the $k_0$-contour integration only the contribution from the residue of the positive-energy pole of the quark $1$ propagator. When all quark pole contributions are included in the $k_0$-contour integration this leads to a coupled set of CST equations, depicted diagrammatically in Fig.\,\ref{fig}. For the heavy and heavy-light systems the 1CSE is a good approximation~\cite{Leitao:2016bqq}, as it retains the most important properties of the complete set of CST equations, i.e. manifest covariance, cluster separability, and the correct one-body and nonrelativistic limits. It is also a good approximation for equal-mass particles, as long as the bound-state mass is large and of the order of the sum of the quark masses. However, a property the 1CSE does not maintain, in general, is charge-conjugation symmetry. Therefore, states calculated with the 1CSE are not expected to be C-parity eigenstates. In principle, this problem is easily remedied by using the set of two-channel CST equations inside the dashed rectangle of Fig.\,\ref{fig} instead.

The 1CSE reads
\begin{equation}
\Gamma_{1CS} (\hat{p}_1,p_2)= - \int \frac{d^3k}{(2\pi)^3} \frac{m_1}{E_{1k}} \sum_K V_K(\hat{p}_1,\hat{k}_1) \Theta_{1}^{K(\mu)}  
\frac{m_1+\hat{\slashed{k}}_1}{2m_1} \Gamma_{1CS}(\hat{k}_1,k_2)
\frac{m_2+\slashed{k}_2}{m_2^2-k_2^2-i\epsilon}\Theta^K_{2(\mu)} \, ,
\label{eq:1CS}
\end{equation}
where $\Theta_i^{K(\mu)}={\bf 1}_i, \gamma^5_i,$ or $\gamma_i^\mu$;  $V_K(\hat{p}_1,\hat{k}_1)$ describes the momentum dependence of the kernel labelled $K$, $m_i$ is the mass of quark $i$, and $E_{ik}\equiv (m_i^2+{\bf k}^2)^{1/2}$. A ``$\hat{\phantom{p}}$'' over a four-momentum indicates that it is on-mass-shell.

The kernel employed in our calculations with the 1CSE consists of a covariant generalization of the linear (L) confining potential used in \cite{Leitao:2014}, a color Coulomb (Coul), and a constant (C) interaction:
\begin{equation}
{\cal V}\equiv
\sum_K V_K  \Theta_{1}^{K(\mu)} \otimes \Theta^K_{2(\mu)}\,= \left[ (1-y) \left({\bf 1}_1\otimes {\bf 1}_2 + \gamma^5_1 \otimes \gamma^5_2 \right) - y\, \gamma^\mu_1 \otimes \gamma_{\mu 2} \right]V_\mathrm{L}  -\gamma^\mu_1 \otimes \gamma_{\mu 2} \left[ V_\mathrm{Coul}+V_\mathrm{C} \right]\,.
\label{eq:kernel}
\end{equation}
The mixing parameter $y$  allows to dial between a scalar-plus-pseudoscalar structure, which preserves chiral symmetry as shown in~\cite{CSTpi-pi}, and a vector structure, while leaving the nonrelativistic limit unchanged. The precise Lorentz structure of the confining interaction is not known, and by fitting the $y$ parameter from the mesonic mass spectra, further information can be gained. Early results favored pure scalar-plus-pseudoscalar confinement, and therefore we set $y=0$ in this work. The momentum-dependent structures of the interaction kernel are
\begin{equation}
V_\mathrm{L}(p,k)  = -8\sigma \pi\left[\left(\frac{1}{q^4}-\frac{1}{\Lambda^4+q^4}\right)-\frac{E_{1p}}{m_1}(2\pi)^3 \delta^3 (\mathbf{q})\int \frac{d^3 k'}{(2\pi)^3}\frac{m_1}{E_{1k'}}\left(\frac{1}{q'^4}-\frac{1}{\Lambda^4+q'^4}\right)\right],
\end{equation}
\begin{equation}
V_\mathrm{Coul}(p,k)  = -4 \pi \alpha_s \left(\frac{1}{q^2}-\frac{1}{q^2-\Lambda^2}\right), \quad
V_\mathrm{C}(p,k) = (2\pi)^3\frac{E_{1k}}{m_1} C \delta^3 (\mathbf{q})\,,
\label{eq:V}
\end{equation}
where $q^{(\prime)}=p-k^{(\prime)}$.

The three coupling strengths, $\sigma$, $\alpha_s$, and $C$, are free parameters of the model. An analysis of the asymptotic behaviour for large momenta  $k$ shows that we need to regularize the kernel. We use Pauli-Villars regularization for both the linear and the Coulomb parts, which yields one additional parameter, the cut-off parameter $\Lambda$. The results turn out not to be very sensitive to the choice of $\Lambda$ and we set $\Lambda=2 m_1$.

Next we expand both the projector and the propagator of Eq.\,(\ref{eq:1CS}) in terms of $u^\rho$-spinors ($\rho=\pm$), defined as 
\begin{equation}
 u_i^{+}(p,\lambda)=\sqrt{\frac{E_{i p}+m_i}{2m_i}} 
 \left(\begin{array}{c}
   1\\
   \frac{\mathbf{\sigma}\cdot \mathbf{p}}{E_{i p}+m_i}
 \end{array}\right)\otimes\chi_\lambda (\hat{p}),
\qquad u_i^{-}(p,\lambda)=\sqrt{\frac{E_{i p}+m_i}{2m_i}} 
 \left(\begin{array}{c}
   -\frac{\mathbf{\sigma}\cdot \mathbf{p}}{E_{i p}+m_i}\\1
 \end{array}\right)\otimes\chi_\lambda (\hat{p})\,,
\label{eq:spinors}
\end{equation}
 where $\chi_\lambda$ are two-component spinors. Introducing the notation
\begin{equation}
\Theta_{i,\lambda\lambda'}^{\rho\rho' K(\mu) }({\bf p},{\bf k}) \equiv
\bar{u}_i^\rho({\bf p},\lambda) \Theta^{K(\mu)} u_i^{\rho'}({\bf k},\lambda'), \qquad \Gamma^{+\rho'}_{\lambda\lambda'} (p) \equiv \bar{u}_1^+({\bf p},\lambda) \Gamma(p) u_2^{\rho'}({\bf p},\lambda'),
\label{eq:Theta,Gamma}
\end{equation}
for the spinor matrix elements of the interaction vertices and of the vertex function, respectively, we obtain
\begin{align}
\Gamma^{+\rho'}_{\lambda\lambda'} (p)
= 
- \int \frac{d^3k}{(2\pi)^3} \frac{m_1}{E_{1k}}  
\frac{ m_2}{E_{2k}}\sum_{\rho\lambda_1 \lambda_2}\sum_K V_K(p,k) \Theta_{1,\lambda\lambda_1}^{++K(\mu)}({\bf p},{\bf k})
\Gamma^{+\rho}_{\lambda_1\lambda_2} (k) 
 \frac{\rho}{E_{2k}-\rho k_{20}}  \Theta_{2,\lambda_2\lambda'K(\mu)}^{\rho\rho'}({\bf k},{\bf p}) \, .
\label{eq:1CSE3}
\end{align}
Multiplying  Eq.~(\ref{eq:1CSE3}) from the left by  $\bar{u}_1^+({\bf p},\lambda)$ and from the right by $u_2^{\rho'}({\bf p},\lambda')$  yields
\begin{multline}
\rho'(E_{2p}-\rho' p_{20}) \sqrt{\frac{m_1 m_2}{E_{1p}E_{2p}} }
\rho'\frac{\Gamma^{+\rho'}_{\lambda\lambda'} (p)}{E_{2p}-\rho' p_{20} }
= 
- \int \frac{d^3k}{(2\pi)^3} \sqrt{\frac{m_1 m_2}{E_{1k}E_{2k}} } \sqrt{\frac{m_1 m_2}{E_{1p}E_{2p}} }\\
\times \sum_{\rho\lambda_1 \lambda_2}\sum_K V_K(p,k) 
\Theta_{1,\lambda\lambda_1}^{++K(\mu)}({\bf p},{\bf k})\sqrt{\frac{m_1 m_2}{E_{1k}E_{2k}} }
\Gamma^{+\rho}_{\lambda_1\lambda_2} (k) 
 \frac{\rho}{E_{2k}-\rho k_{20}}  \Theta_{2,\lambda_2\lambda'K(\mu)}^{\rho\rho'}({\bf k},{\bf p}) \, .
\label{eq:1CSE4}
\end{multline}
Introducing the CST wave functions when quark 1 is on-shell,
\begin{equation}
 \Psi^{+\rho}_{1,\lambda_1\lambda_2}(k) \equiv \sqrt{\frac{m_1m_2}{E_{1k}E_{2k}}}\frac{\rho}{E_{2k}-\rho (E_{1k}-\mu)} \Gamma^{+\rho}_{\lambda_1\lambda_2}(k),
\label{eq:wfs1}
\end{equation}
we can finally cast Eq.\,(\ref{eq:1CS}) into the form
\begin{multline}
(\rho' E_{2p}-E_{1p}+\mu) \Psi^{+\rho'}_{1,\lambda\lambda'} (p)
= 
- \int \frac{d^3k}{(2\pi)^3}  N(p,k)
\sum_{\rho\lambda_1 \lambda_2}\sum_K V_K(p,k) 
\Theta_{1,\lambda\lambda_1}^{++K(\mu)}({\bf p},{\bf k}) \\
\times \Psi^{+\rho}_{1,\lambda_1\lambda_2} (k) 
 \Theta_{2,\lambda_2\lambda'K(\mu)}^{\rho\rho'}({\bf k},{\bf p})
\label{eq:1CSE6}
\end{multline}
where $N(p,k) = m_1 m_2/\sqrt{E_{1k}E_{2k}E_{1p}E_{2p} }$.
The CST wave functions can be written in terms of two-component spinors $\chi_\lambda$ and  $K_j^\rho(\hat{\bf p})$ operators, which are $2\times 2$ matrices that depend on the total angular momentum and the parity of the meson under study (in this work we consider $J^P=0^{\pm}, 1^{\pm}$),
\begin{equation}
\Psi^{+\rho}_{1,\lambda\lambda'} (p)=\sum_j \psi_j^\rho(p) \chi^\dagger_\lambda(\hat{\bf p})\, K_j^\rho(\hat{\bf p}) \, \chi_{\lambda'}(\hat{\bf p}).
\label{eq:PSI}
\end{equation}

\begin{table}[tbh]
\caption{Wave function components for the mesons considered in this work.}
\begin{center}
\begin{tabular}{l||cc|cc||cc|cc}
\hline
$J^P$ & $K_1^-(\hat{\bf p})$ &Wave& $K_2^-(\hat{\bf p})$ &Wave& $K_1^+(\hat{\bf p})$ &Wave&  $K_2^+(\hat{\bf p})$ &Wave\\
\hline
$0^-$      & ${\bf 1}$ & $S$&- &- & ${\bm \sigma}\cdot{\bf \hat p}$ & $P$ &- &- \\
$0^+$     & ${\bm \sigma}\cdot{\bf \hat p}$ & $P$& - & - &  ${\bf 1}$ & $S$ &- &-\\
\hline
$1^-$      & $\sigxi$  & $S$& $\frac{1}{\sqrt{2}}\left(3 \xip\, \sigp-\sigxi\right)$  & $D$ & $\sqrt{3}\xip$ & $P_s$ & $\sqrt{\frac32}\left(\sigxi\, \sigp-\xip\right) $ & $P_t$ \\
$1^+$      & $\sqrt{3}\xip$  & $P_s$&  $\sqrt{\frac32}\left(\sigxi\, \sigp-\xip\right)$  & $P_t$ & $\sigxi$ & $S$ & $\frac{1}{\sqrt{2}}\left(3 \xip\, \sigp-\sigxi\right)$ & $D$ \\
\hline
\end{tabular}
\end{center}
\label{tab:KK1}
\end{table}

Table \ref{tab:KK1} lists the $K_j^\rho(\hat{\bf p})$ used in this work. 
The main advantage of using this basis for the wave function is that it explicitly displays its orbital-angular-momentum content and thus enables us to determine the spectroscopic identity of our solutions, which is indispensable when comparing to the measured states. Our wave functions contain relativistic components not present in nonrelativistic solutions. For instance, the $S$-waves of our pseudoscalar states couple to small $P$-waves (with opposite intrinsic parity) that vanish in the nonrelativistic limit, whereas, for vector mesons, coupled $S$- and $D$-waves are accompanied by relativistic spin-singlet  and spin-triplet  $P$-waves, denoted $P_s$ and $P_t$, respectively.

\section{Results and Discussion}
\label{sec:3}
In this work we present two models: model P1 was fitted to the masses of pseudoscalar states only, whereas model PSV1 was fitted to the masses of pseudoscalar, scalar, and vector mesons.\,The parameters of the models are listed in Table~\ref{tab:parameters}. The constituent quark masses were first determined in preliminary calculations and then held fixed in the final fits of $\sigma$, $\alpha_s$, and $C$. 
\begin{table}[h!]
\caption{Parameters of models P1 and PSV1. Both models use the quark masses $m_b=4.892$ GeV, $m_c=1.600$ GeV, $m_s=0.448$ GeV, and $m_u=m_d\equiv m_q=0.346$ GeV. Shown are also the number of states used in the fits, $N_\mathrm{fit}$, and the respective rms differences between the model predictions and the data.}
\centering
\label{tab:parameters}       % Give a unique label
%\begin{tabular}{l||lllll}
\begin{tabular}{c|ccccc}
\hline\noalign{\smallskip}
Model & $\sigma$ [GeV$^2$] & $\alpha_s$ & $C$ [GeV] &  $N_\mathrm{fit}$ & rms difference [GeV]\\[3pt]
\tableheadseprule\noalign{\smallskip}
P1      & 0.2493 & 0.3643 & 0.3491 & 9& 0.036\\
PSV1 	& 0.2247 & 0.3614 & 0.3377 & 25 &0.031\\
\hline
\end{tabular}
\end{table}

\begin{table}[h!]
\centering
\caption{Comparison of the masses of experimental meson states (Exp) with at with least one $b$ or $c$ quark and  $J^P=0^\pm,1^\pm$ to the theoretical predictions of models P1 and PSV1. The $\triangle$ and $\square$ symbols indicate the states used in the fits of models P1 and PSV1, respectively. The states with no symbol assigned are pure predictions. The experimental values for axial-vector mesons in $b\bar{b}$ and $c\bar{c}$ are averages over the two possible charge-conjugation parities. All masses are given in units of GeV. There is weak evidence (at the level of 1.8 $\sigma$) that the $\Upsilon(1D)$ ($10.15$ GeV, marked with "?")  has been seen \cite{CLEO,BABAR}.}
\label{tab:masses}       % Give a unique label
% For LaTeX tables you can use
\begin{tabular}{l|lll|lll|lll|lll}
&\multicolumn{3}{|c|}{$J^P=0^-$}    &   \multicolumn{3}{|c|}{$J^P=1^-$} & \multicolumn{3}{|c|}{$J^P=0^+$}  & \multicolumn{3}{|c}{$J^P=1^+$}  \\
\hline
 % $J^P=0^-$ & &  &   $J^P=1^-$ & &  &   $J^P=0^+$&  &  &  $J^P=1^+$ & & \\ \hline 
&Exp & P1 & PSV1 & Exp & P1 & PSV1& Exp & P1 & PSV1 & Exp & P1 & PSV1\\  \hline
& 9.398$^{\triangle, \square}$ & 9.386 & 9.415 & 9.460$^{\square}$ & 9.470 & 9.487 & 9.859$^{\square}$ & 9.856 &9.850 & 9.896 & 9.886 & 9.875 \\
& 9.999$^{\triangle, \square}$ & 9.982 & 9.968 & 10.02$^{\square}$ & 10.02 & 10.00 & 10.23$^{\square}$ & 10.25 & 10.22 & 10.26 & 9.890 & 9.879 \\
$b\bar{b}$& 10.30 & 10.37 & 10.33 & 10.15(?) & 10.16 & 10.13 & - & 10.57 & 10.52 & 10.51 & 10.27 & 10.24 \\
& - & 10.68 & 10.63 & 10.36$^{\square}$ & 10.40 & 10.35 & - & 10.86 & 10.80 & - & 10.28 & 10.24 \\
& - & 10.96 & 10.89 & - & 10.49 & 10.44 & - & 11.13 & 11.03 & - & 10.60 & 10.54 \\
& - & 11.28 & 11.16 & 10.58$^{\square}$ & 10.71 & 10.65 & - & 11.48 & 11.32 & - & 10.60 & 10.54 \\ \hline
&6.275$^{\triangle, \square}$ & 6.302 & 6.319 & - & 6.394 & 6.397 & - & 6.745 & 6.730 & - &6.777 & 6.757  \\
$b\bar{c}$& 6.842 & 6.888 & 6.865 & - & 6.941 & 6.912 & - & 7.161 & 7.121 & - & 6.777 & 6.758 \\
& - & 7.293 & 7.246 & - & 7.057 & 7.019 & - & 7.505 & 7.445 & - & 7.191 & 7.146 \\\hline
& 5.367$^{\triangle, \square}$ & 5.362 & 5.367 & 5.415$^{\square}$ & 5.442 & 5.436 & - & 5.784 & 5.763 & 5.829 & 5.796 & 5.770 \\
$b\bar{s}$& - & 5.938 & 5.910 & - & 5.993 & 5.957 & - & 6.208 & 6.163 & - & 5.811 & 5.785 \\
& - & 6.349 & 6.297 & - & 6.093 & 6.051 & - & 6.559 & 6.495 & - & 6.234 & 6.184 \\ \hline
& 5.279$^{\triangle, \square}$ & 5.288 & 5.293 & 5.325$^{\square}$ & 5.366 & 5.360 & - & 5.709 & 5.688 &5.726 & 5.716 & 5.690 \\
$b\bar{q}$& - & 5.864 & 5.835 & - & 5.918 & 5.882 & - & 6.132 & 6.087 & - &5.735 & 5.708 \\
& - & 6.274 & 6.221 & - & 6.017 & 5.974 & - & 6.483 & 6.418 & - &6.157 & 6.106 \\ \hline
& 2.984$^{\triangle, \square}$ & 3.009 & 3.030 & 3.097$^{\square}$ & 3.110 & 3.120 & 3.415$^{\square}$& 3.424 & 3.424 & 3.518 & 3.461 & 3.454 \\
$c\bar{c}$& 3.639$^{\triangle, \square}$ & 3.647 & 3.627 & 3.686$^{\square}$ & 3.702 & 3.677 & 3.918 & 3.930 & 3.894 & - & 3.474 & 3.465 \\
& - & 4.123 & 4.073 & 3.773$^{\square}$ & 3.784 & 3.756 & - & 4.355 & 4.291 & - & 3.950 & 3.911 \\ \hline
& 1.968$^{\triangle, \square}$ & 1.944 & 1.966 & 2.112$^{\square}$ & 2.107 & 2.109 & 2.318$^{\square}$ & 2.399 & 2.396 & 2.459 & 2.434 & 2.422 \\
$c\bar{s}$& - & 2.612 & 2.591 & - & 2.697 & 2.667 & - & 2.910 & 2.872 & 2.535 & 2.458 & 2.444 \\
& - & 3.100 & 3.048 & - & 2.769 & 2.737 & - & 3.340 & 3.274 & - & 2.934 & 2.893 \\ \hline
& 1.867$^{\triangle, \square}$ & 1.858 & 1.881 & 2.009$^{\square}$ & 2.029 & 2.030 & 2.318$^{\square}$ & 2.319 &2.316 & 2.421 & 2.351 & 2.339 \\
$c\bar{q}$& - & 2.529 & 2.507 & - & 2.617 & 2.587 & - & 2.828 & 2.790 & - & 2.377 & 2.362 \\
& - & 3.016 & 2.964 & - & 2.687 & 2.655 & - & 3.257 & 3.191 & - & 2.852 & 2.810 \\ \hline
\end{tabular}
% Or use
%\vspace*{5cm}  % with the correct table height
\end{table}
Our results for $b\bar{b}$, $b\bar{c}$, $b\bar{s}$, $b\bar{q}$, $c\bar{c}$, $c\bar{s}$, $c\bar{q}$ states ($q$ stands for a $u$ or a $d$ quark) are given in Table~\ref{tab:masses}. The calculated masses are very close to the experimental data, with an rms difference of roughly $30$ MeV for both models P1 and PSV1. These differences are comparable with those reported in Ref.~\cite{Spence} and, more recently, in Ref.~\cite{YangLi}, and give us confidence that our approach is appropriate for the systems under study. 

The parameters of the two models are very similar, as Table~\ref{tab:parameters} shows.
This is remarkable, because pseudoscalar states (the only ones used in the fit of model P1) are essentially pure $S$-waves and are not sensitive to spin-orbit and tensor forces. Moreover, in pseudoscalar mesons the spin-spin forces act only in spin-singlet states and therefore cannot be separated from the central forces. This means that no information about the spin-dependent interactions was used to constrain the kernel of model P1. On the other hand, the scalar and vector states to which model PSV1 was fitted \emph{are} sensitive to the spin-dependent interactions and do constrain its parameters. The fact that both fits give essentially the same parameters and rms differences to the data means that the spin-dependent interactions are \emph{correctly predicted} by the covariant kernel of model P1. 

\begin{figure}[h!]
\centering
\includegraphics[width=1.0\textwidth]{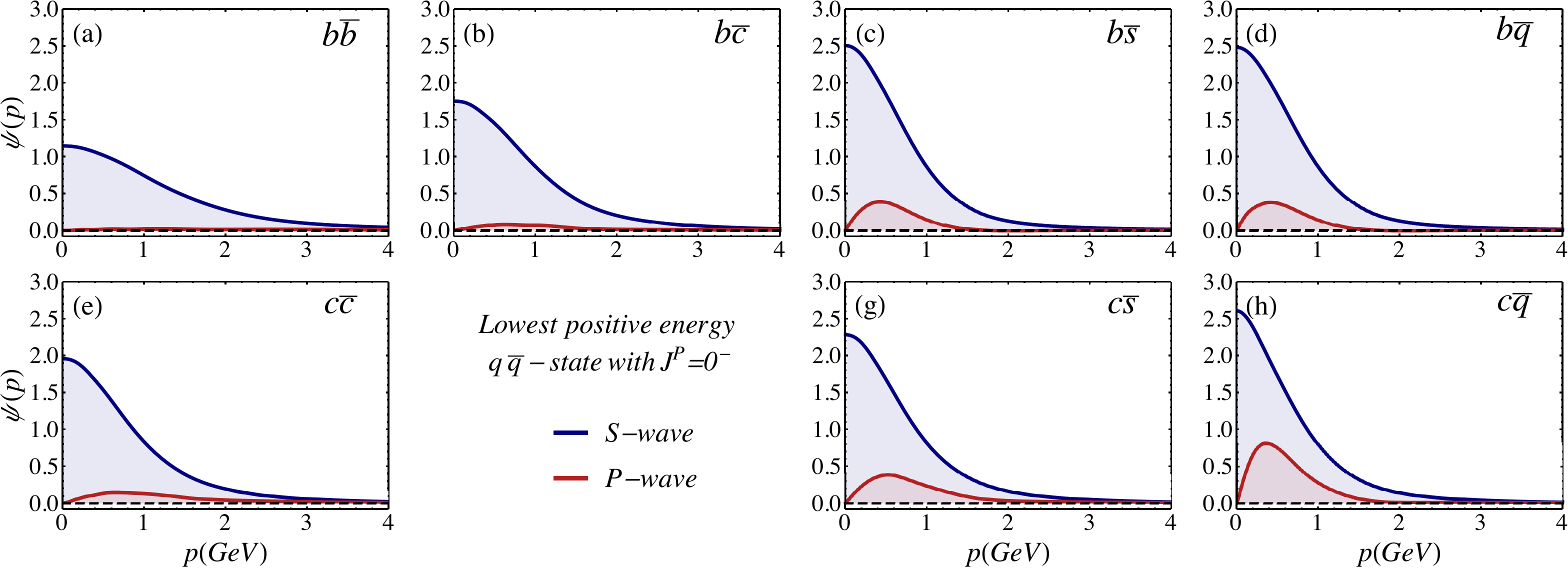}
\caption{Model P1 ground-state wave functions for pseudoscalar mesons ($J^P=0^-$).}
\label{fig:groundps}       % Give a unique label
\end{figure}

\begin{figure}[H]
\centering
\includegraphics[width=1.0\textwidth]{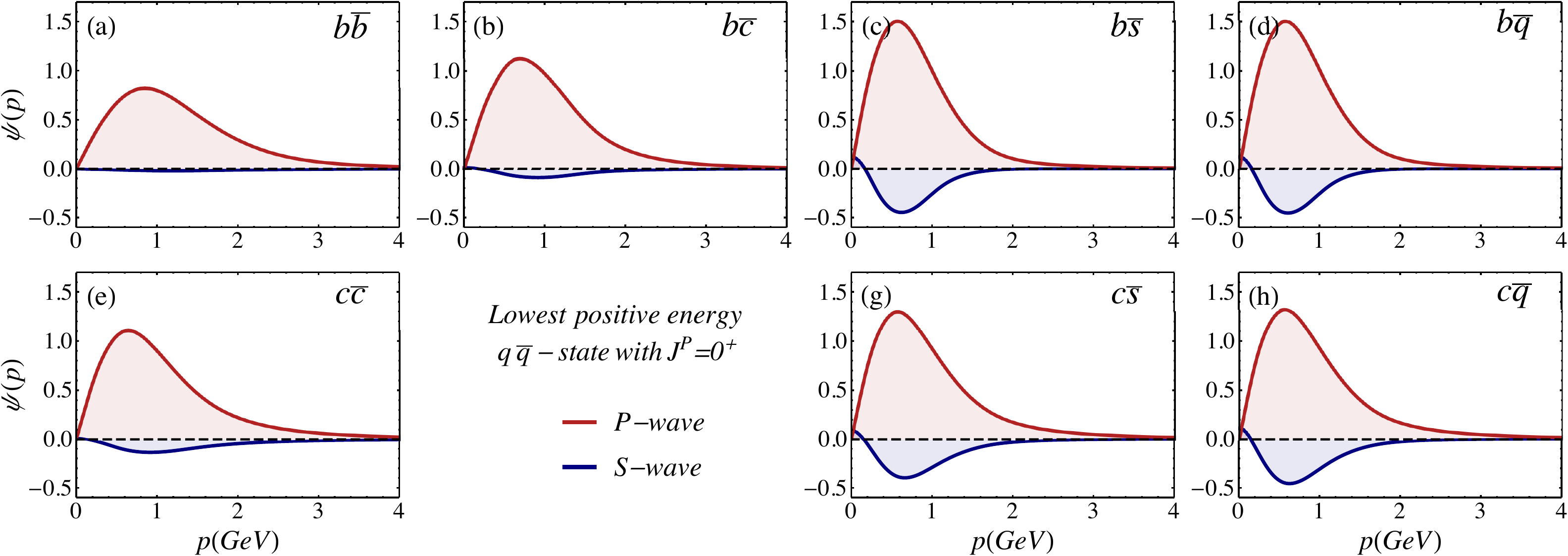}
\caption{Model P1 ground-state wave functions for scalar mesons ($J^P=0^+$).}
\label{fig:grounds}       % Give a unique label
\end{figure}

\begin{figure}[h!]
\centering
\includegraphics[width=1.0\textwidth]{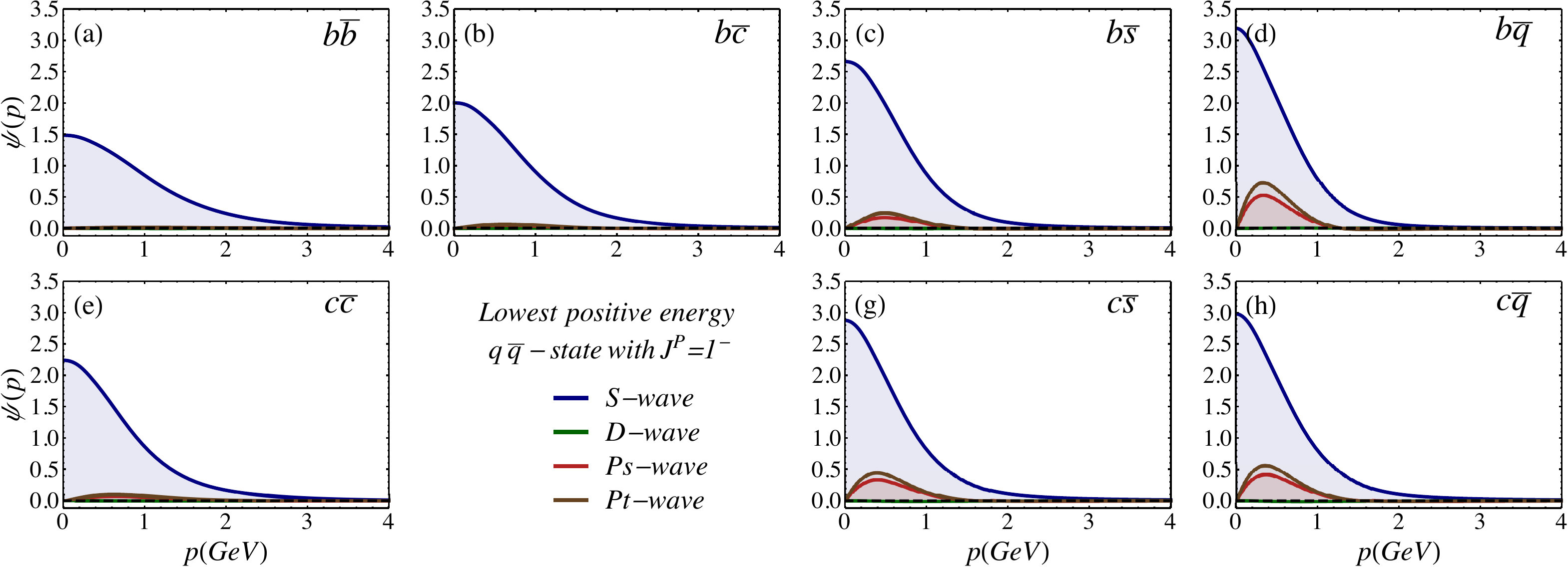}
\caption{Model P1 ground-state wave functions for vector mesons ($J^P=1^-$).}
\label{fig:groundv}       % Give a unique label
\end{figure}

\begin{figure}[h!]
\centering
\includegraphics[width=1.0\textwidth]{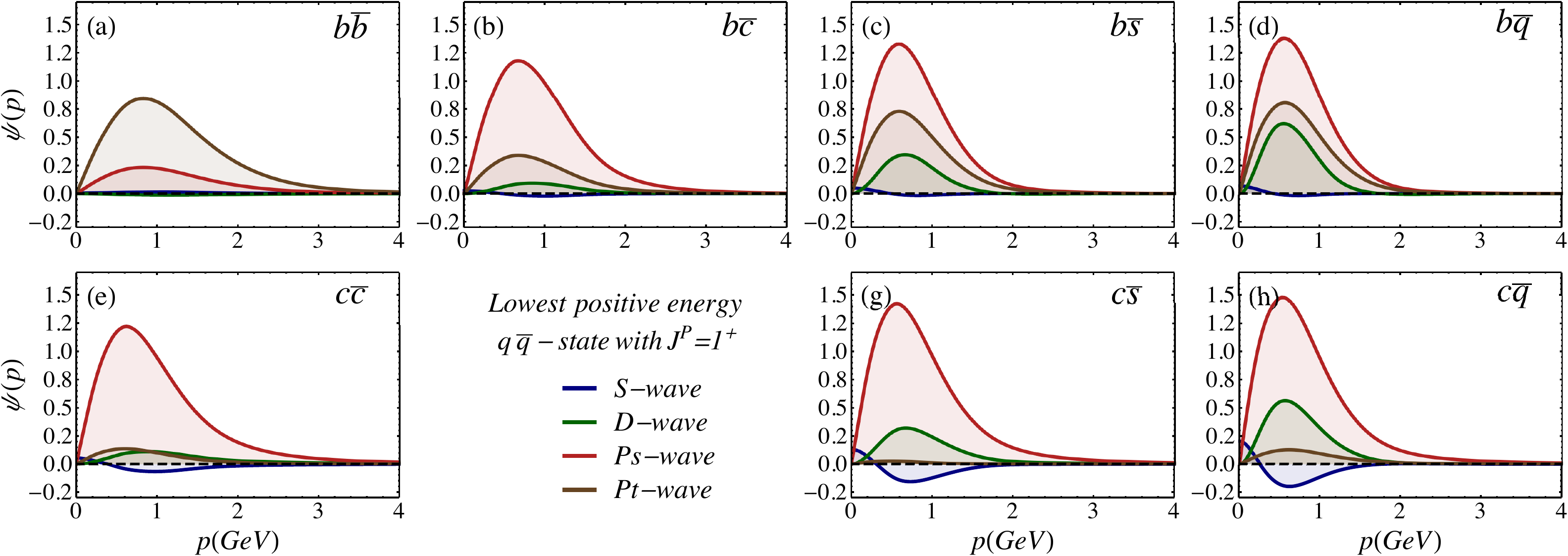}
\caption{Model P1 ground-state predictions for axial-vector mesons $J^P=1^+$.}
\label{fig:groundav}       % Give a unique label
\end{figure}

We have also calculated the wave functions needed to study the structure of mesons and to calculate form factors and decay rates. In Figures~\ref{fig:groundps}--\ref{fig:groundav}, the ground-state wave function components $\psi(p)$, defined as in Eq.\,(\ref{eq:PSI}) and calculated with model P1, are depicted for pseudoscalar, scalar, vector and axial-vector mesons.\,These components are normalized as
\vspace{-0.4cm}
\begin{align}
\int dp \, p^2 \left[\psi^2_S(p)+\psi^2_D(p) \right]=1\,, \qquad \text{for } J^P=0^{\pm}\,,\\
\int dp \, p^2 \left[\psi^2_S(p)+\psi^2_D(p)+\psi^2_{P_s}(p)+\psi^2_{P_t}(p) \right]=1\,, \qquad \text{for } J^P=1^{\pm}\,.
\end{align}
Tables~\ref{tab:p}--\ref{tab:av} list the probabilities of the different wave function components. As expected, the relativistic components are tiny in heavy mesons, but become larger as the constituent quark masses decrease, reaching almost 10\% for $c\bar{q}$ mesons, the lightest systems considered here. 
The momentum-space wave functions tend to be more spread out for heavier states, and as one decreases the mass of the lighter quark, the wave functions concentrate more in the region of low momenta.

\begin{table}[h!]
\caption{Probabilities of $S$- and $P$-wave components for pseudoscalar mesons ($J^P=0^-$), left panel, and  scalar mesons ($J^P=0^+$), right panel.}
\centering
\label{tab:p}       % Give a unique label
\begin{tabular}{c|c|ccc}
\hline\noalign{\smallskip}
Meson & Fig.~2 &  $S$-wave (\%) & $P$-wave (\%) \\
\hline\vspace{-5pt}\\
$b\bar{b}$&(a) & 99.5& 0.528 \\
$b\bar{c}$&(b) &99.4&0.586 \\
$b\bar{s}$&(c) & 96.8&3.19 \\
$b\bar{q}$&(d) &95.7&4.34 \\
\hline\vspace{-5pt}\\
$c\bar{c}$&(e) &96.3&3.71 \\
$c\bar{s}$&(g) &92.0&8.04 \\
$c\bar{q}$&(h) &90.8&9.15 \\
\hline
\end{tabular}
\qquad\qquad
\begin{tabular}{c|c|ccc}
\hline\noalign{\smallskip}
Meson & Fig.~3 &  $S$-wave (\%) & $P$-wave (\%) \\
\hline\vspace{-5pt}\\
$b\bar{b}$&(a) & 0.203&99.8 \\
$b\bar{c}$&(b) &0.855&99.1 \\
$b\bar{s}$&(c) & 6.77&93.2 \\
$b\bar{q}$&(d) &6.04&94.0 \\
\hline\vspace{-5pt}\\
$c\bar{c}$&(e) &2.97&97.0 \\
$c\bar{s}$&(g) &6.38&93.6 \\
$c\bar{q}$&(h) &9.24&90.8\\
\hline
\end{tabular}
\end{table}

%\begin{table}[h!]
%\caption{Probabilities of $S$- and $P$-wave components for scalar mesons ($J^P=0^+$).}
%\centering
%\label{tab:s}       % Give a unique label
%\begin{tabular}{c|c|ccc}
%\hline\noalign{\smallskip}
%Meson & Fig.~3 &  $S$-wave (\%) & $P$-wave (\%) \\
%\hline\vspace{-5pt}\\
%$b\bar{b}$&(a) & 0.203&99.8 \\
%$b\bar{c}$&(b) &0.855&99.1 \\
%$b\bar{s}$&(c) & 6.77&93.2 \\
%$b\bar{q}$&(d) &6.04&94.0 \\
%\hline\vspace{-5pt}\\
%$c\bar{c}$&(e) &2.97&97.0 \\
%$c\bar{s}$&(g) &6.38&93.6 \\
%$c\bar{q}$&(h) &9.24&90.8\\
%\hline
%\end{tabular}
%\end{table}

\begin{table}[h!]
\caption{Probabilities of $S$-, $D$-, $P_s$- and $P_t$-wave components for vector mesons ($J^P=1^-$).}
\centering
\label{tab:v}       % Give a unique label
\begin{tabular}{c c cccc}
\hline\noalign{\smallskip}
Meson & Fig.4 &$S$-wave (\%)&$D$-wave (\%)&$P_s$-wave (\%)&$P_t$-wave  (\%)\\
\hline\vspace{-5pt}\\
$b\bar{b}$&(a) &99.9&0.0130&0.0186&0.0477\\
$b\bar{c}$&(b) &99.6&0.0118&0.144&0.196\\
$b\bar{s}$&(c) &97.6&0.00862&0.858&1.49\\
$b\bar{q}$&(d) &93.5&0.00783&2.30&4.22 \\
\hline\vspace{-5pt}\\
$c\bar{c}$&(e) &98.9&0.0124&0.397&0.731\\
$c\bar{s}$&(g) &95.3&0.0309&1.87&2.78 \\
$c\bar{q}$&(h) &94.0&0.0354&2.38&3.61 \\
\hline
\end{tabular}
\end{table}

\begin{table}[h!]
\caption{Probabilities of $S$-, $D$-, $P_s$- and $P_t$-wave components for axial-vector mesons ($J^P=1^+$).}
\centering
\label{tab:av}       % Give a unique label
\begin{tabular}{c c cccc}
\hline\noalign{\smallskip}
Meson & Fig.~5 &$S$-wave (\%)&$D$-wave (\%)&$P_s$-wave (\%)&$P_t$-wave  (\%)\\
\hline\vspace{-5pt}\\
$b\bar{b}$&(a) &0.0559&0.0399&7.54&92.4\\
$b\bar{c}$&(b) &0.0262&0.670&89.3&9.97\\
$b\bar{s}$&(c) &0.0134&5.93&71.9&22.1\\
$b\bar{q}$&(d) &0.0111&7.34&69.7&23.0 \\
\hline\vspace{-5pt}\\
$c\bar{c}$&(e) &0.578&1.66&97.0&0.750\\
$c\bar{s}$&(g) &1.14&6.39&92.4&0.0954 \\
$c\bar{q}$&(h) &1.15&7.90&90.5&0.397\\
\hline
\end{tabular}
\end{table}

\section{Summary and Outlook}
\label{sec:4}
In this work we report on recent progress in our studies of heavy and heavy-light mesons in the framework of the CST. Our kernel consists of covariant versions of a linear confining potential with Lorentz-scalar coupling, in combination with Lorentz-vector one-gluon-exchange plus a constant interaction kernel.  When fitting a small number of global adjustable parameters we obtain an accurate description of the mass spectrum, which gives us confidence that our goal of a global description of all $q\bar q$ mesons may indeed be feasible. 

From the observation that a fit to the masses of mesons whose wave functions are essentially pure S-waves leads to almost exactly the same parameters as another fit to meson states whose orbital angular momenta are not restricted, we conclude that the covariance of our interaction kernel accurately predicts the spin-dependent quark-antiquark interactions. 

We also compared the orbital-angular-momentum components of the various meson wave functions and found that they behave as expected: the lighter the constituent quark masses of the quark-antiquark system, the larger become the probabilities of the wave function components of purely relativistic origin.

For the near future we plan to include also heavy and heavy-light tensor mesons in our calculations with the one-channel spectator equation. The next step is then to extend our model to the light-meson sector, which requires the use of the more complicated four-channel equations. In addition, with the covariant meson wave functions we will compute other observables such as decay rates, and study the structure of mesons in more detail by calculating electroweak elastic  and transition form factors.

\begin{acknowledgements}
This work was supported by the Portuguese Funda\c{c}\~ao para a Ci\^encia e a Tecnologia (FCT) under contracts SFRH/BD/92637/2013, SFRH/BPD/100578/2014, and UID/FIS/0777/2013.
\end{acknowledgements}

% BibTeX users please use
%\bibliographystyle{spbasic}


\end{document}